\documentclass[iop]{emulateapj}

\usepackage{graphicx}
\usepackage{dcolumn}
\usepackage{bm}
\usepackage{epsf}

\def\agt{\mathrel{\raise.3ex\hbox{$>$}\mkern-14mu\lower0.6ex\hbox{$\sim$}}}
\def\alt{\mathrel{\raise.3ex\hbox{$<$}\mkern-14mu\lower0.6ex\hbox{$\sim$}}}

\newcommand{\beq}{\begin{equation}}
\newcommand{\eeq}{\end{equation}}
\newcommand{\beqn}{\begin{eqnarray}}
\newcommand{\eeqn}{\end{eqnarray}}
\newcommand{\pa}{\partial}

\newcommand{\varep}{\varepsilon}

\begin{document}

\title{Afterglow of binary neutron star merger}

\author{
Masaru Shibata\altaffilmark{1},
Yudai Suwa\altaffilmark{1},
Kenta Kiuchi\altaffilmark{1},
and Kunihito Ioka\altaffilmark{2}
}

\altaffiltext{1}{Yukawa Institute for Theoretical Physics, Kyoto
  University, Oiwake-cho, Kitashirakawa, Sakyo-ku, Kyoto 606-8502,
  Japan}
\altaffiltext{2}{KEK Theory Center and the Graduate University for
  Advanced Studies, Oho, Tsukuba 305-0801, Japan}

\begin{abstract}
The merger of two neutron stars results often in a rapidly and
differentially rotating hypermassive neutron star (HMNS). We show by
numerical-relativity simulation that the magnetic-field profile around
such HMNS is dynamically varied during its subsequent evolution, and
as a result, electromagnetic radiation with a large luminosity $\sim
0.1 B^2 R^3 \Omega$ is emitted with baryon ($B$, $R$, and $\Omega$ are
poloidal magnetic-field strength at stellar surface, stellar radius,
and angular velocity of a HMNS). The predicted luminosity of
electromagnetic radiation, which is primarily emitted along the
magnetic-dipole direction, is $\sim 10^{47} (B/10^{13}~{\rm
  G})^2(R/10~{\rm km})^3(\Omega/10^4~{\rm rad/s})$~ergs/s, that is
comparable to the luminosity of quasars.
\end{abstract}


\section{Introduction}
Coalescence of binary neutron stars (BNS) is one of the most promising
sources for next-generation kilo-meter size gravitational-wave
detectors such as advanced LIGO, advanced VIRGO, and LCGT. Statistical
studies have suggested that the detection rate of gravitational waves
emitted by BNS will be $\sim 1$--100 per year~\citep{Rate}. Typical
signal-to-noise ratio for such detection will be $\sim 10$ or less.
Thus, it will be crucial for the detection of gravitational waves to
find electromagnetic counterparts to the gravitational-wave signals.
Short-hard gamma-ray bursts (SGRB) have been inferred to accompany
with the BNS merger~\citep{GRB-BNS-1,GRB-BNS-2}.  However, this
hypothesis relies on many uncertain assumptions; e.g., high
magnetic-field strength or efficient pair annihilation of
neutrino-antineutrino. In this article, we give a conservative
estimate for the strength of electromagnetic signals based on a
numerical-relativity simulation and show that a strong electromagnetic
signal will indeed accompany with the BNS merger.

BNS evolves due to gravitational radiation reaction and eventually
merges. After the merger sets in, there are two possible fates~(e.g.,
\cite{STU2,KSST}): If the total mass $M$ is larger than a critical
mass $M_c$, a black hole will be formed, while a hypermassive neutron
star (HMNS) will be formed for $M < M_c$.  The value of $M_c$ depends
strongly on the equation of state (EOS) of neutron stars, but the
latest discovery of a high-mass neutron star with mass $1.97 \pm
0.04M_{\odot}$~\citep{twosolar} indicates that the EOS is stiff and
$M_c$ may be larger than the typical total mass of the BNS $\sim
2.7M_{\odot}$~\citep{Stairs}.  This indicates that HMNS is the likely
outcome for many BNS mergers, at least temporarily~\citep{hotoke}.

Neutron stars in nature have a strong magnetic field with typical
field strength at the stellar surface $10^{11}$--$10^{13}$~G.  One of
the neutron stars in BNS often has field strength smaller than this
typical value as $\sim 10^{10}$--$10^{11}$~G \citep{Lorimer}, probably
because of the accretion history of the first-formed neutron star
during the formation of the second one. However, at least the second
one is likely to have the typical magnetic-field strength.

It is reasonable to believe that each neutron star in the inspiral
phase (before the merger sets in) has an approximately dipole magnetic
field as in the isolated one. During the late inspiral phase and
formation of a HMNS in the merger phase, its magnetic-field profile
will be modified due to magnetohydrodynamics (MHD) processes. However,
in zeroth approximation, it would be safe to suppose that the
dipole field is dominant. Due to this reason, we consider the
evolution of a HMNS with dipole magnetic fields in the following.

One of the most important properties of HMNS is that it is {\em
  rapidly} and {\em differentially} rotating~\citep{STU2}. The
numerical simulations have shown that the typical angular velocity at
its center is $\Omega \agt 10^4$~rad/s, much larger than that of
ordinary pulsars, while at equatorial surface it is $\sim
10^3$~rad/s\footnote{We note that even for rigidly rotating neutron
  stars, rotating magnetic field lines with a high degree of
  differential rotation are produced in the vicinity of the neutron
  stars because of the presence of a light cylinder close to their
  equatorial surface at $\sim c/\Omega=30(\Omega/10^4~{\rm
    rad/s})^{-1}$~km with $c$ being the speed of light.}. Because of
the presence of the differential rotation, the winding of magnetic
fields is enhanced: Toroidal magnetic-field strength $B_T$ in HMNS
increases linearly with time ($t$) in the presence of seed poloidal
(cylindrically radial) magnetic fields $B_P$ ($B_T$ increases as $\sim
B_P \Omega t$). The increase of the magnetic-field strength results in
the increase of magnetic pressure.  Because only dilute matter is
present in the surface of HMNS, Alfv\'en waves are likely to
propagate near the rotational axis with $\varpi \alt 10$~km, where
$\varpi=\sqrt{x^2+y^2}$, transporting electromagnetic energy generated
in HMNS along the rotational axis; tower-type outflow is
driven. As far as the HMNS is alive and the rapid rotation is present,
the amplification of the toroidal magnetic field continues via the
winding effect. Then, the electromagnetic energy should increase
approximately as $\dot E_B \sim B_P^2 V \Omega$ as described
in~\cite{meier} where $V$ is an effective volume for which the
amplification occurs\footnote{We note that the luminosity of winds by
  the magnetocentrifugal effect~\citep{BP} has the same order of
  magnitude, but with the vertical-dominant dipole fields considered
  here, this effect is not dominant.  The magnetic dipole radiation
  could also play an important role as in ordinary
  pulsars~\citep{dip9,dip0,dip1} because its luminosity $\propto B^2
  R^6 \Omega^4/c^3$~\citep{ST83} may be comparable to the luminosity
  of Eq.~(\ref{eq1}) for HMNS with $R\Omega/c > 0.1$. However, the
  property of electromagnetic wave emission (e.g., emission direction)
  would be different from that we consider in this paper.}. If the
amplification efficiently occurs near the rotation axis with $\varpi
\alt 10$~km, $V$ is approximately $\sim (\alpha R)^3$ where $R$ is the
equatorial stellar radius $\sim 10$~km and $\alpha$ is a constant of
$O(0.1)$. For a typical HMNS formed after the merger
\beqn 
\dot E_B &\sim &10^{46} B_{13}^2~R_6^3~\Omega_4
\alpha_{0.1}^3~{\rm ergs/s}, \label{eq1}
\eeqn
where $B_{13}=B_P/10^{13}$~G, $R_6=R/10$~km,
$\Omega_4=\Omega/10^4$~rad/s, and $\alpha_{0.1}=\alpha/0.1$. We
suppose a relatively high magnetic-field strength because it is likely
to be amplified by compression that occurs during the
merger~\citep{AEI}.  Thus, the luminosity of the electromagnetic
radiation will be as high as that of quasars for a typical field
strength of a progenitor neutron star $B_P \sim 10^{12}$~G and of a
resulting HMNS $B_P \sim 10^{13}$~G. If a substantial fraction of this
generated electromagnetic energy is converted to electromagnetic
radiation (as suggested, e.g., in~\cite{NP11}), the merger event may
be detected by telescopes as the electromagnetic signals.

\section{Numerical simulation}
Motivated by the fact mentioned above, we performed MHD simulation for
a HMNS in general relativity. As the initial condition, we prepare a
rapidly and differentially rotating HMNS in axisymmetric equilibrium
as in~\cite{SDLSS}: We constructed a HMNS model with the following
piecewise polytropic EOS: $P=P_{\rm cold}=K_1\rho^{\Gamma_1}$ for
$\rho \leq \rho_{\rm nuc}$ and $P_{\rm cold}=K_2\rho^{\Gamma_2}$ for
$\rho \geq \rho_{\rm nuc}$. Here, $P$ and $\rho$ are the pressure and
rest-mass density. We set $\Gamma_1=1.3$, $\Gamma_2=2.75$, $K_1=5.16
\times 10^{14}$~cgs, $K_2=K_1\rho_{\rm nuc}^{\Gamma_1-\Gamma_2}$, and
$\rho_{\rm nuc}=1.8 \times 10^{14}~{\rm g/cm^3}$. With this EOS, the
maximum gravitational mass (rest mass) is
$2.01M_{\odot}~(2.32M_{\odot})$ for spherical neutron stars and
$2.27M_{\odot}~(2.60M_{\odot})$ for rigidly rotating neutron
stars. These are similar values to those in realistic stiff EOS (e.g.,
~\cite{EOS} for a review).  We prepare a HMNS with the following
physical parameters: gravitational mass $M=2.65M_{\odot}$, baryon rest
mass $M_b=2.96M_{\odot}$, maximum density $\rho_{\rm max}=9.0 \times
10^{14}~{\rm g/cm^3}$, angular momentum $J=0.82GM^2/c$, ratio of polar
to equatorial radius 0.3, central rotation period $P_c =0.202$~ms, and
rotation period at the equatorial surface $5.4P_c$. Here, $G$ is the
gravitational constant.  The rotation law is specified in the same way
as in~\cite{BSS} with the differential rotation parameter $\hat
A=0.8$. This HMNS is similar to that found in the BNS merger
simulation of~\cite{STU2,KSST} performed in nuclear-theory-based
EOS. In the evolution, we employ an EOS of the form $P=P_{\rm
  cold}+(\Gamma_1-1)\rho (\varep-\varep_{\rm cold})$ where $\varep$ is
the specific internal energy and $\varep_{\rm cold}$ is its cold part
determined from $P_{\rm cold}$.

Poloidal magnetic fields, for which the toroidal component of the
vector potential has the form $A_{\varphi}=A_0\varpi_0\varpi^2/(r^2 +
\varpi_0^2)^{3/2}$, are superimposed on this initial condition.  The
magnetic field in the inertial frame is given by
$B^i=\epsilon^{ijk}\pa_j A_k$ where $\epsilon_{ijk}$ is the completely
anti-symmetric tensor.  Here, $r^2=\varpi^2+z^2$, and $(\varpi_0,
A_0)$ are constants: $\varpi_0$ is chosen to be $5R_e/3$--$20R_e/3$
where $R_e(=12.1~{\rm km})$ is the coordinate radius on the equatorial
plane. We found that the electromagnetic luminosity shown below
depends weakly on this parameter. In the following we show the results
for $\varpi_0=10R_e/3$ and $20R_e/3$.  $A_0$ determines the field
strength for which we give the maximum magnetic-field strength $B_{\rm
  max} \approx 10^{13}$--$4\times 10^{14}$~G. Here, the magnetic-field
strength is defined by $B=\sqrt{b_{\mu}b^{\mu}}$ where $b^{\mu}$ is
the 4-vector of the magnetic field in the frame comoving with the
fluid. With such strength, the magnetic pressure in the HMNS is much
smaller than the matter pressure at its center, and thus, the density
profile of the HMNS (except for its surface) is not significantly
modified by the magnetic-field effect. Because of the presence of
differential rotation, the magnetic fields may be amplified in the
exponential manner due to the magnetorotational instability
(MRI)~\citep{MRI}.  However, the wavelength for the fastest growing
mode is very short ($\sim 10^2$--$10^4$~cm) in the present setting and
it is not possible to resolve in numerical simulation. We here do not
pay attention to the MRI but only to the winding effect.  In the
presence of the MRI effects, the electromagnetic energy will be
increased more rapidly and the luminosity of electromagnetic radiation
may be even enhanced. Thus, this work would determine the lower bound
of the magnetic luminosity, that is, however, quite high.

\begin{figure*}[t]
\includegraphics[width=0.24\textwidth]{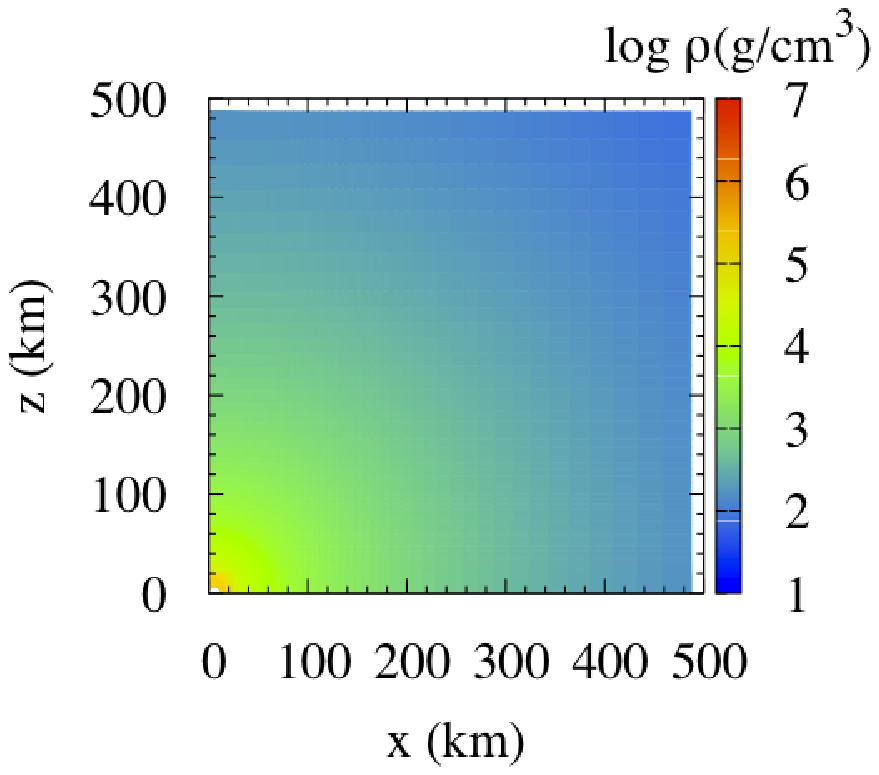}
\includegraphics[width=0.24\textwidth]{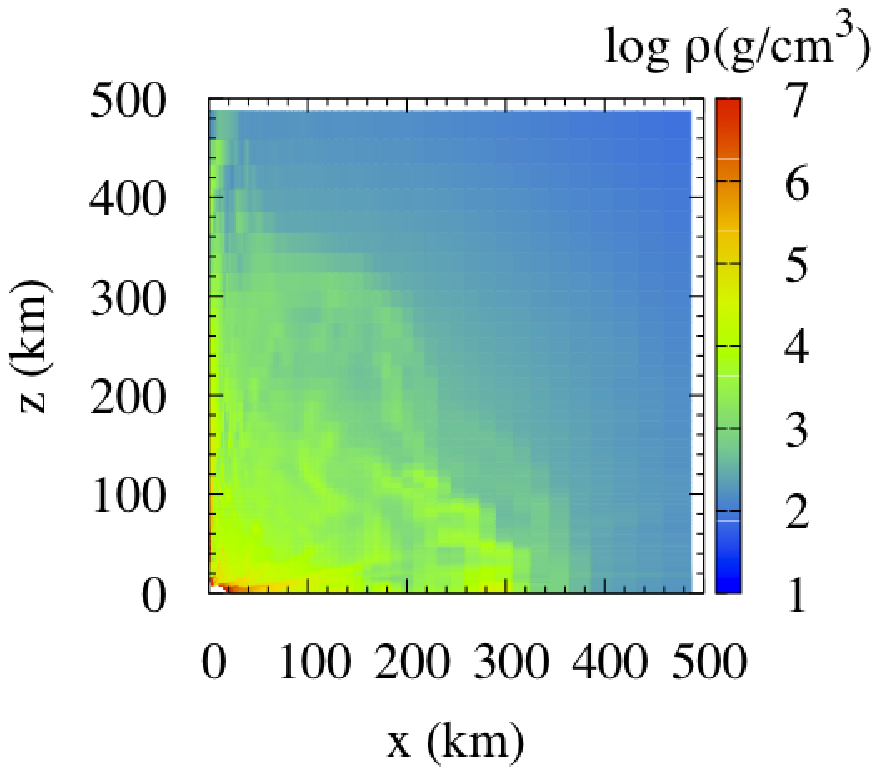}
\includegraphics[width=0.24\textwidth]{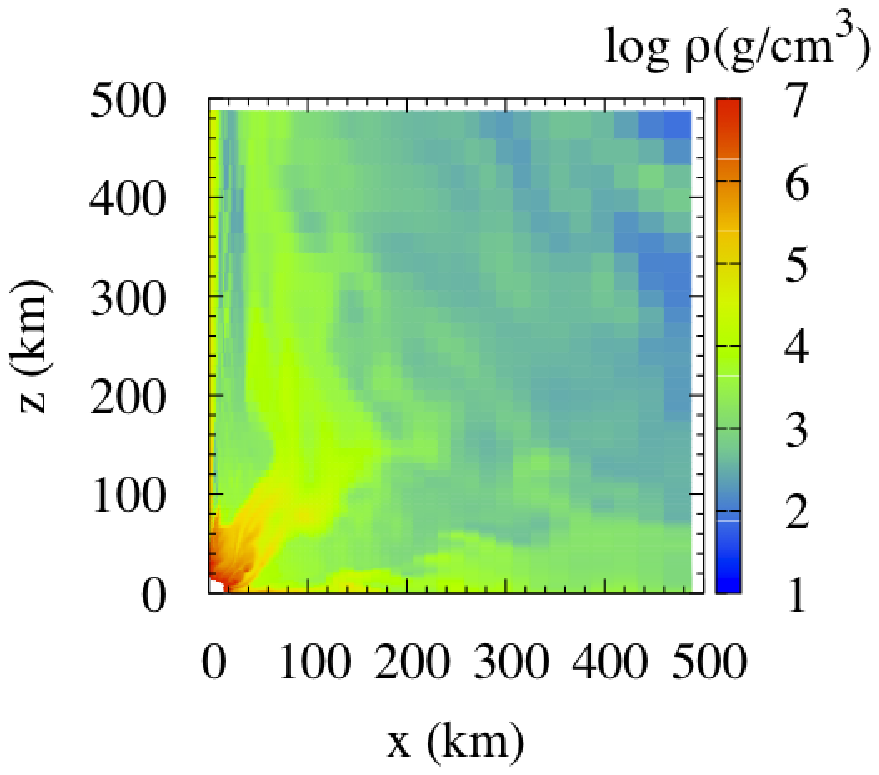}
\includegraphics[width=0.24\textwidth]{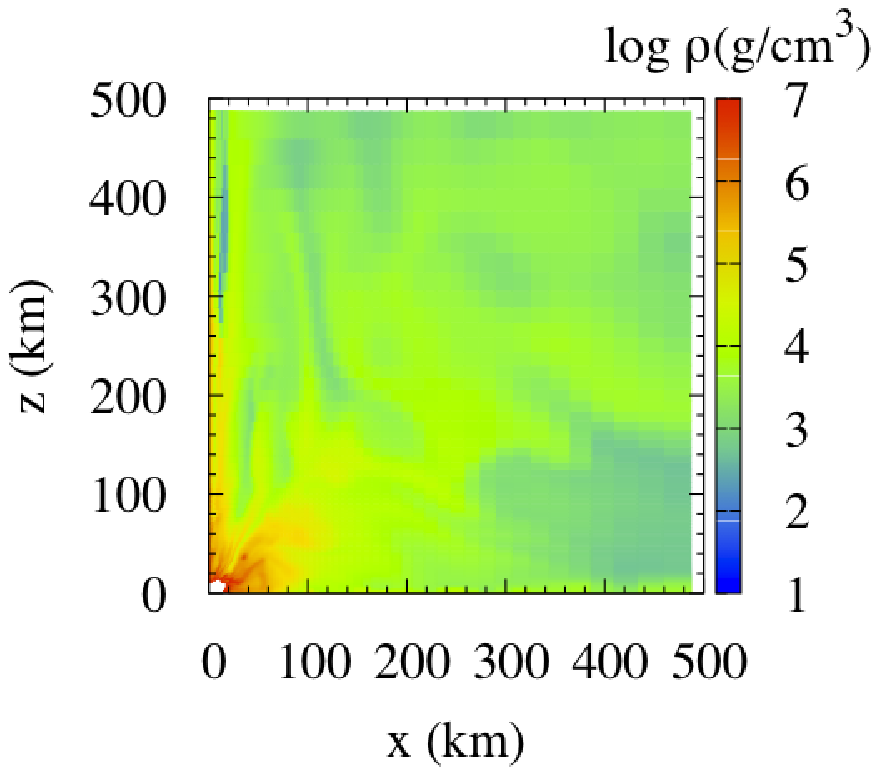}

\includegraphics[width=0.24\textwidth]{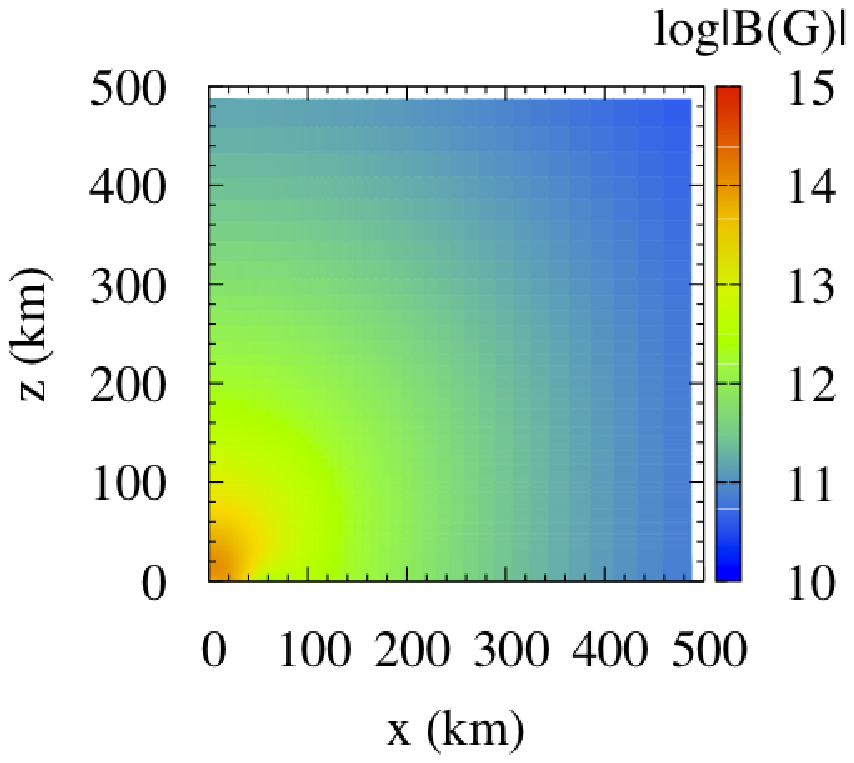}
\includegraphics[width=0.24\textwidth]{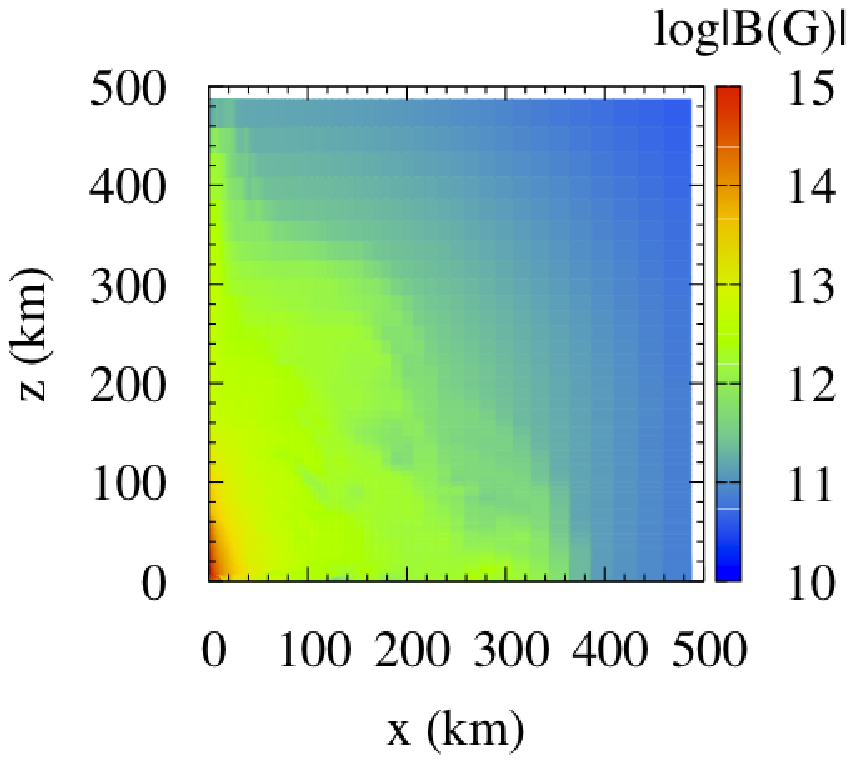}
\includegraphics[width=0.24\textwidth]{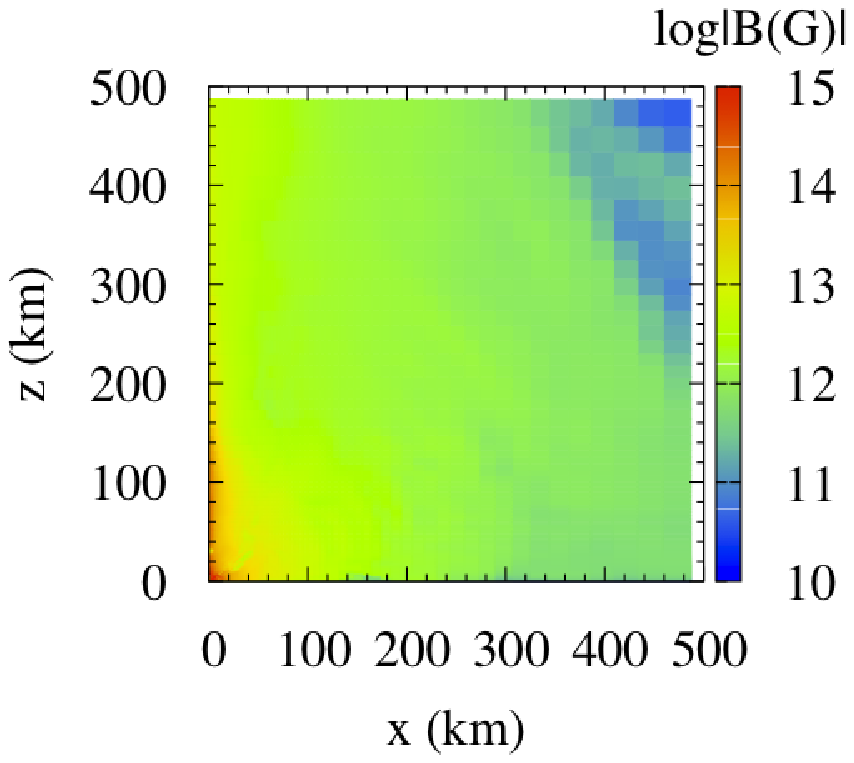}
\includegraphics[width=0.24\textwidth]{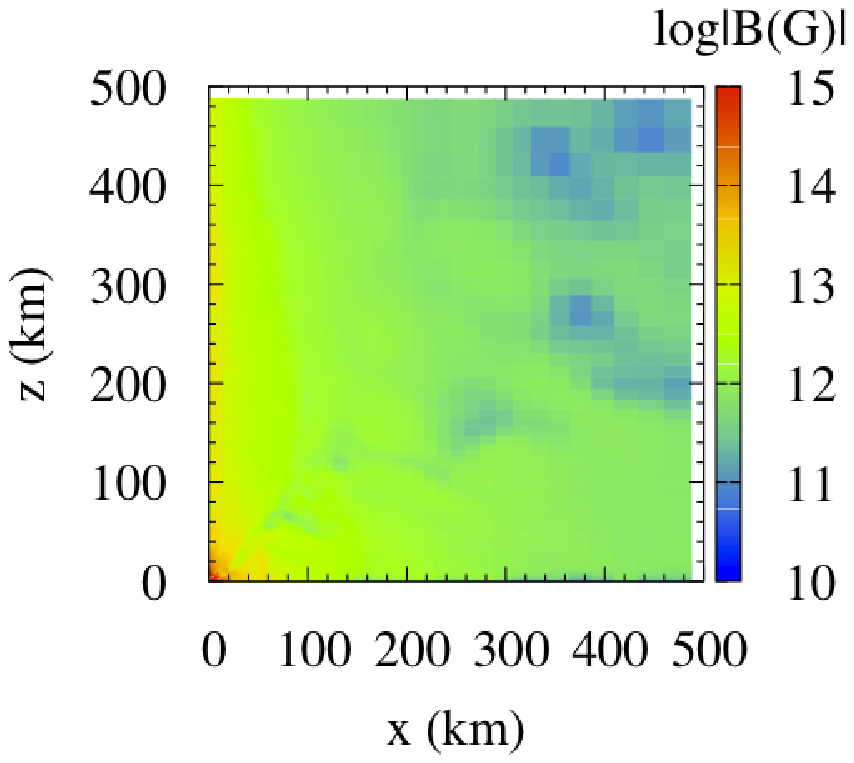}
\caption{Snapshots of density profiles (upper panels) and the
  magnetic-field strength profiles (lower panels) outside HMNS at
  $t=0$, 2.8, 6.2, and 12.7 ms for a model with $B_{\rm max}=1.7
  \times 10^{14}$~G, $\varpi_0=10R_e/3$, and $n=2$. The region with
  $\rho > 10^7~{\rm g/cm^3}$ are shown to be white in the upper
  panels.
\label{FIG1}}
\end{figure*}

MHD simulation is performed assuming that ideal MHD condition holds. A
conservative shock capturing scheme is employed for solving MHD
equations as in~\cite{SS05}: In the present work, a numerical scheme
with third-order accuracy in space and fourth-order accuracy in time
is employed. Einstein's evolution equations are solved in the
fourth-order accuracy in space and time in the so-called BSSN-puncture
formulation~\citep{BSSN-1,BSSN-2,BSSN-3}.

Any conservative scheme in MHD cannot handle vacuum, and hence, we
have to add an atmosphere of small density outside the HMNS.  Because
the matter outside a HMNS formed in a real merger would be dilute, the
density of the atmosphere should be as small as possible to exclude
spurious effects by it. We set the density of the atmosphere as 
\beqn
\rho_{\rm at}=\Big\{
\begin{array}{ll}
f_{\rm at} \rho_{\rm max} & r < 2R_e, \\
f_{\rm at} \rho_{\rm max}(r/2R_e)^{-n} & r \geq 2R_e, \\
\end{array}
\eeqn
where we choose $n=2$ or 2.5.  $f_{\rm at}$ is constant, for which we
typically give $10^{-9}$. We changed the values of $f_{\rm at}$ from
$10^{-10}$ to $10^{-7}$ for $B_{\rm max}=4.2 \times 10^{13}$~G and
found that as far as $B^2/(4\pi \rho_{\rm at}c^2) \alt 1$, our code
works well.  For a large value of $f_{\rm at}$, the evolution of
magnetic fields is substantially affected by the inertia of the
matter. However, with decreasing the value of $f_{\rm at}$ to
$B^2/(4\pi \rho_{\rm at}c^2) \sim 1$, the dependence of magnetic-field
evolution on the atmosphere density becomes weak, and hence, the
effect of the artificial atmosphere does not play a role.  The
velocity of the atmosphere is set to be zero initially.  With this
treatment, the magnetic field is initially modified for $t \alt
P_c$. However, such modification plays a minor role after the winding
effect becomes dominant for the magnetic field amplification. We
always perform simulations for a time much longer than $P_c$, and
focus on the stage for which a quasisteady state is achieved. Thus,
the artificial effect associated with the initial setting does not
matter.

Axisymmetric numerical simulation is performed in the cylindrical
coordinates for MHD and in the Cartesian coordinates for Einstein's
equation part (using the so-called Cartoon method). The details are
described in~\cite{SS05} and its references to which the reader may
refer. Nonuniform grid is prepared as in~\cite{KSY}.  We here impose
the axial symmetry to guarantee a sufficiently high grid resolution
although we can perform a nonaxisymmetric simulation.  In the present
setting, the equatorial coordinate radius of the HMNS, $R_e$, is
covered by 150 uniform grids. Smaller grids with 100 and 120 were also
adopted to check the convergence of the numerical results.  Outer
boundaries along $x$ and $z$ axes are located at $\approx 170R_e
\approx 2000$~km.

Figure \ref{FIG1} plots the evolution of the density profiles and
magnetic-field strength outside the HMNS.  Numerical simulation shows
that the system evolves in the following manner. Because of the
presence of differential rotation, the winding of magnetic fields
proceeds. Toroidal field strength is increased linearly with time, and
the growth rate is high in particular near the rotational axis
($\varpi \alt 10$~km).  Hence, Alfv\'en waves propagate primarily
toward $z$-direction near the rotational axis, and magnetic-field
strength there also increases. After the substantial winding, the
magnetic pressure $B^2/8\pi$ becomes larger than the gravitational
potential energy density near the polar surface, $\sim GM\rho/H$ where
$H(=0.3R_e)$ is the vertical coordinate radius of the HMNS. Then, the
matter of the HMNS in the vicinity of its polar surface is stripped,
and an outflow is driven. Because the magnetic energy density
$B^2/4\pi$ is comparable to or slightly smaller than the rest-mass
density $\sim (H/GM)(B^2/8\pi) > B^2/4\pi c^2$, the outflow is mildly
relativistic with the velocity of order $0.1c$ in the vicinity of the
HMNS.  However, in the region far from the HMNS, the outflow velocity
could be fairly relativistic $\sim 0.9c$ near the rotation axis (see
Fig.~\ref{FIG3}).

\begin{figure*}[t]
\begin{center}
\includegraphics[width=0.45\textwidth]{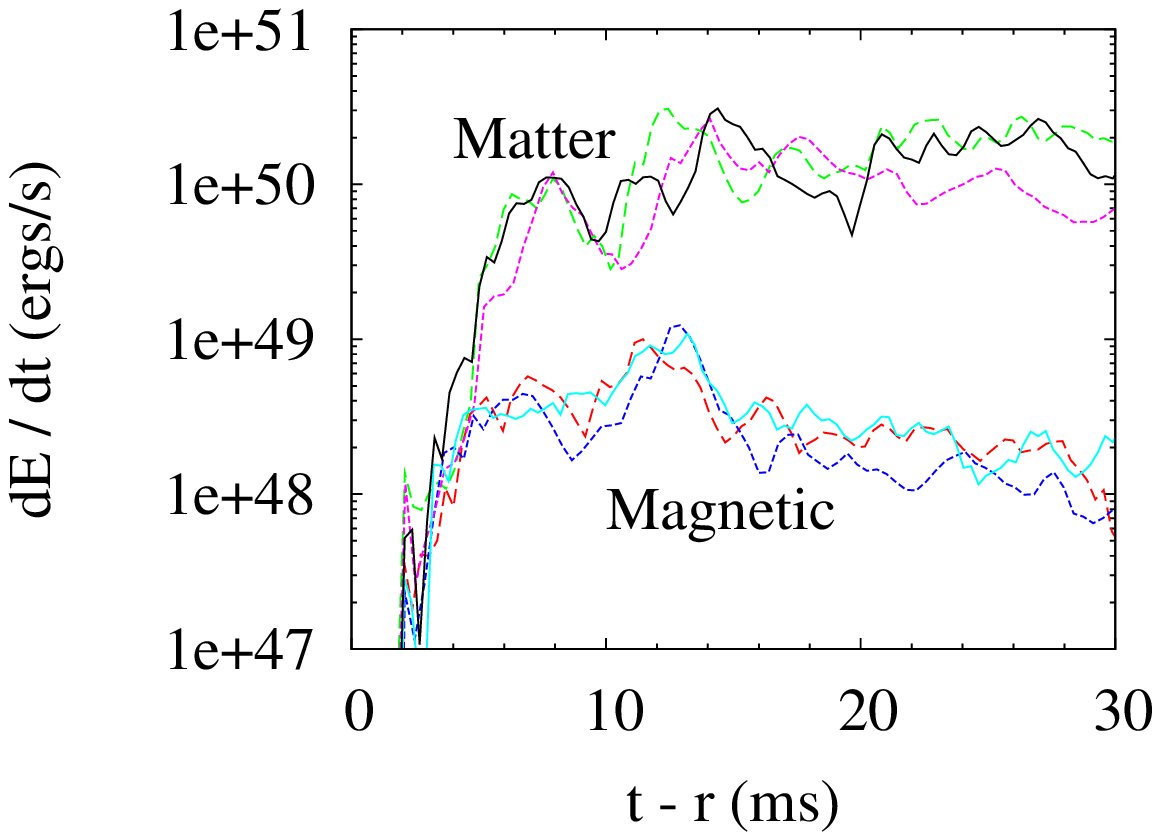}
\includegraphics[width=0.45\textwidth]{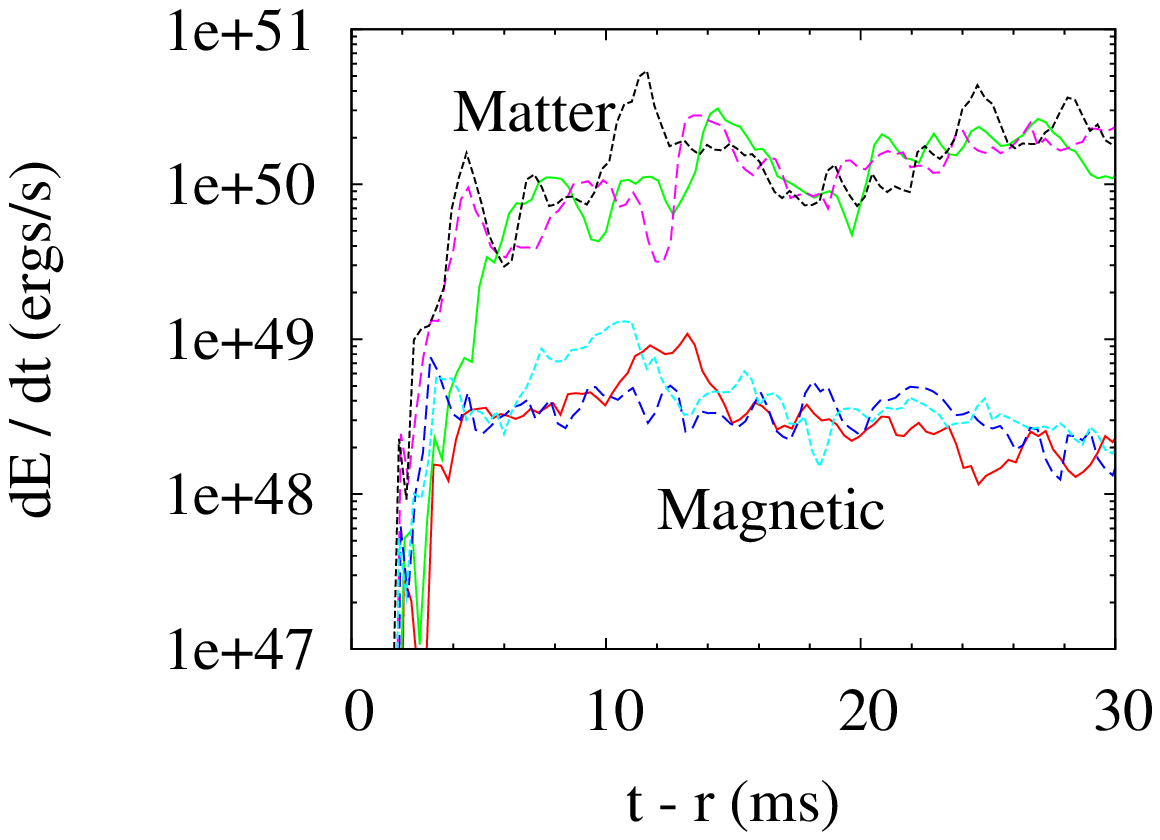}
\includegraphics[width=0.45\textwidth]{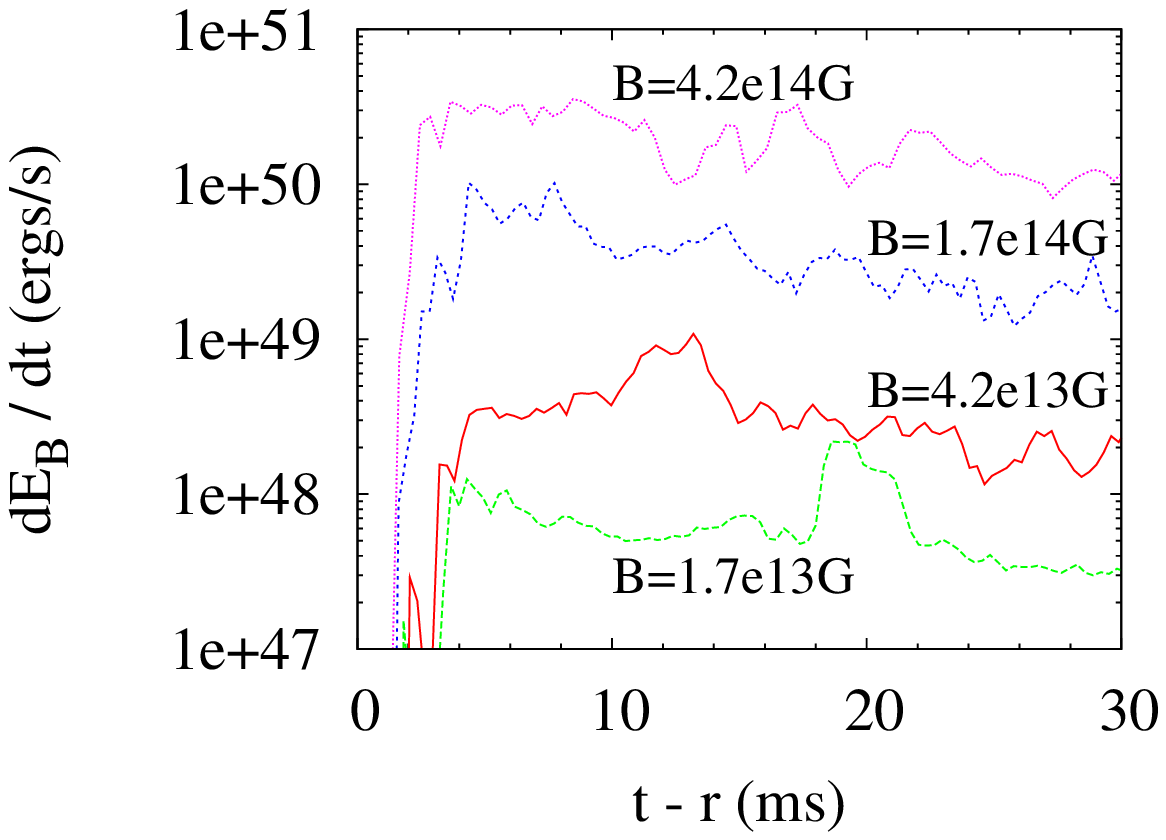}
\includegraphics[width=0.45\textwidth]{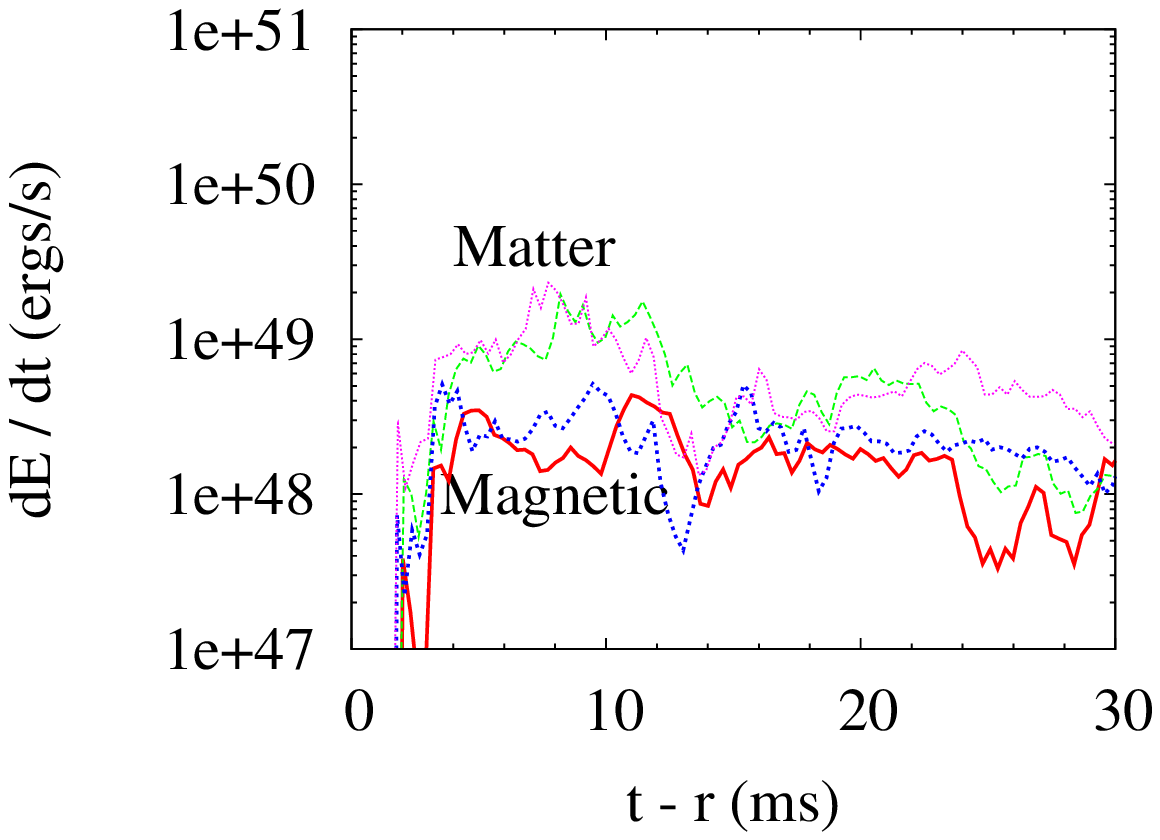}
\end{center}
\caption{ Top-left: Total matter energy (upper curves) and magnetic
  energy (lower curves) ejection rates as functions of time for
  $B_{\rm max}=4.2 \times 10^{13}$~G, $\varpi_0=10R_e/3$, and $n=2$.
  The results with three different grid resolutions are plotted: The
  solid, dashed, and dotted curves show the results with high, middle,
  and low resolutions.  Top-right: The same as the top-left panel but
  for $(\varpi_0,n)=(10R_e/3,2)$ (solid curves), $(10R_e/3, 2.5)$
  (dashed curves), and $(20R_e/3,2)$ (dotted curves).  Bottom-left:
  Evolution of $\dot E_B$ for $B_{\rm max}=1.7 \times 10^{13}$, $4.2
  \times 10^{13}$, $1.7 \times 10^{14}$, and $4.2 \times 10^{14}$~G.
  Bottom-right: The same as the top panels but for the amount
  integrated only for the angle $\Delta\theta=\pi/20$~radian from the
  rotation axis for $(\varpi_0,n)=(10R_e/3,2)$ (solid curves) and
  $(20R_e/3,2)$ (dotted curves).  The thick curves denote the
  electromagnetic luminosity.  For all the figures, $r$ denotes the
  radius for which the surface integral of Eq.~(\ref{dedt}) is carried
  out.
\label{FIG2}}
\end{figure*}

Figure~\ref{FIG2} plots evolution of the matter and electromagnetic
energy ejection rates $\dot E_M$ and $\dot E_B$ defined by
\beqn
\dot E=-\oint_{r={\rm const}} d\Omega \sqrt{-g} T^r_{~t}, 
\label{dedt}
\eeqn
where we substitute the stress energy tensor for the matter and
electromagnetic fields into $T^r_{~~t}$, respectively. $g$ is the
determinant of the spacetime metric. The surface integral is performed
for $r \approx 480$~km. We checked that the luminosity depends only
weakly on the radius of the surface integral.  The top-left and
top-right panels show the evolution of $\dot E_M$ and $\dot E_B$ for
$B_{\rm max}=4.2 \times 10^{13}$~G, and the bottom-left panel show the
evolution of $\dot E_B$ for $B_{\rm max}=1.7 \times 10^{13}$--$4.2
\times 10^{14}$~G.  The bottom-right panel shows the ejection rates
integrated only for the angle $\Delta \theta=\pi/20$ for $B_{\rm
  max}=4.2 \times 10^{13}$~G.  The top-left panel is for
$\varpi_0=10R_e/3$, and $n=2$ with three grid resolutions, and the
top-right panel is for different values of $\varpi_0$ and $n$. The
top-left and top-right panels show that irrespective of the grid
resolution, $\varpi_0$, and $n$, the total ejection rates are \beqn
\dot E_M &\sim & 10^{48} B_{13}^2 R_6^3 \Omega_4~{\rm ergs/s}, \\ \dot
E_B &\sim & 10^{47} B_{13}^2 R_6^3 \Omega_4~{\rm ergs/s}.  \eeqn The
bottom-left panel indeed shows that the scaling relation with respect
to the magnetic-field strength holds (the same scaling also holds for
$\dot E_M$).  Another important point is that these ejection rates do
not significantly vary in time.  Thus, a quasisteady outflow is
driven.

The value of $\dot E_B$ agrees approximately with the prediction of
Eq.~(\ref{eq1}), implying that the scenario described in Sec.~1 is
correct. The ratio of $\dot E_M/\dot E_B$ is of order $2c^2 H/ G M
\sim 10$. This agrees approximately with the value required for the
mass stripping.

Comparison among top-left, top-right, and bottom-right panels of
Fig.~\ref{FIG2} shows that the electromagnetic energy is mainly
emitted in the direction near the rotation axis. By contrast, the
matter energy is emitted in a fairly isotropic manner. Along the
rotational axis, $\dot E_M/\dot E_B$ is of order unity ($\sim
1$--10). This is the reason that the outflow along the rotation axis
can be mildly relativistic.


\begin{figure*}[t]
\begin{center}
\includegraphics[width=0.45\textwidth]{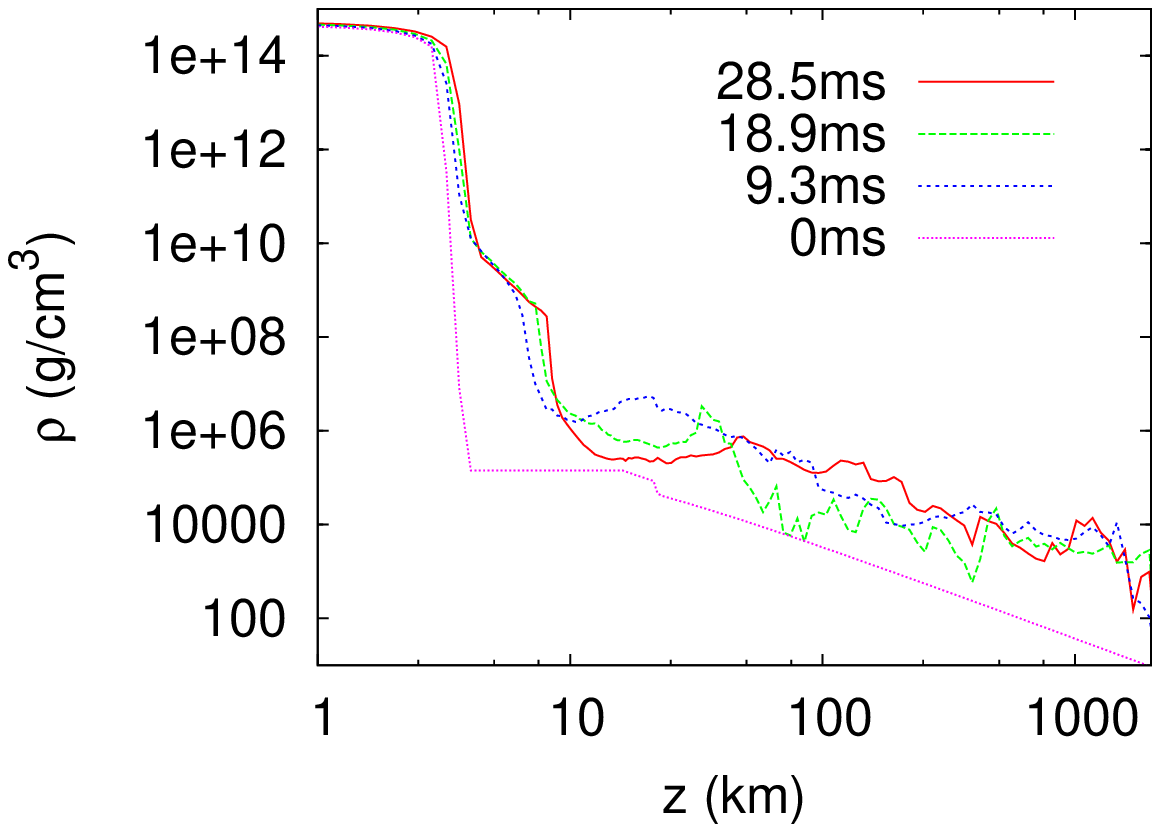}
\includegraphics[width=0.45\textwidth]{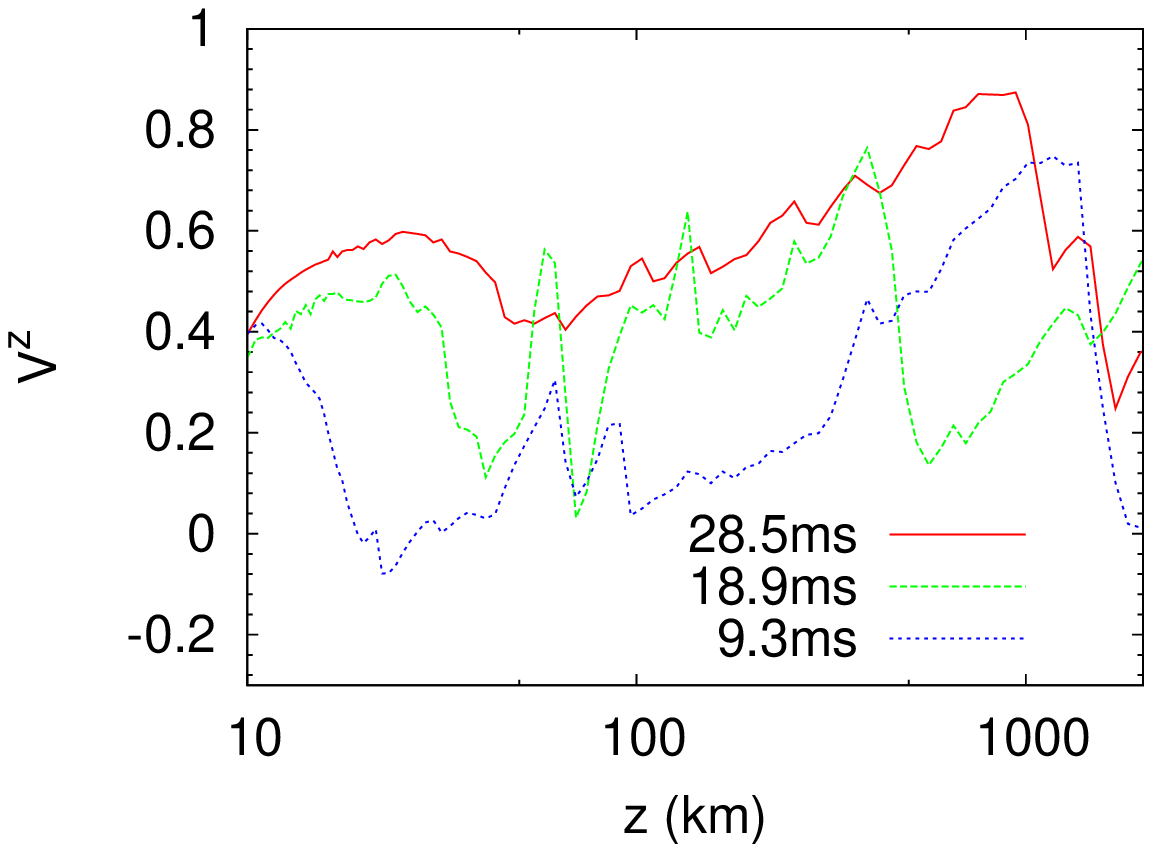}
\end{center}
\caption{Snapshots of density (left panel) and velocity profiles
  (right panel) along the rotation axis for $B_{\rm max}=4.2 \times
  10^{13}$~G, $\varpi_0=10R_e/3$, and $n=2$. At $t=0$, the density is
  determined by the atmosphere and decreases in the form $\rho \propto
  r^{-2}$ for the large radius.
\label{FIG3}}
\end{figure*}

The amount of angular momentum loss by the matter ejection and
electromagnetic radiation in the time duration $\Delta t=P_c$ is much
smaller than the total angular momentum because of our choice of
magnetic-field strength $\alt 3\times 10^{14}$~G. This implies that
the matter and electromagnetic waves are continuously ejected, and a
quasisteady outflow is formed. Figure~\ref{FIG3} plots the density and
velocity profiles along the rotation axis for $B_{\rm max}=4.2 \times
10^{13}$~G, $\varpi_0=10R_e/3$, and $n=2$. The density profile (left
panel) indeed shows that the averaged density does not change
significantly with time. The density decreases with the radius. In
these examples, the power-law index is roughly $n_{\rho} \sim 1.5$--2
($\rho \propto r^{-n_{\rho}}$), but this number depends on the initial
setting for $B_{\rm max}$ and $\varpi_0$ and varies with time.  The
velocity profile (right panel of Fig.~\ref{FIG3}) shows that the
outflow is mildly relativistic. The profile varies in a short time
scale. The maximum velocity is $\sim 0.9c$ as mentioned above: This
maximum depends weakly on the setting of the atmosphere; for the lower
atmosphere density, the maximum speed is larger.  The averaged
magnitude of the outflow velocity in time is $\sim 0.4$--$0.6c$ (which
also depends weakly on the setting of the atmosphere).  Because the
corresponding Lorentz factor of the jet is $\Gamma_j\lesssim 2$, the
relativistic beaming effect (i.e., observable viewing angle is
$\lesssim 1/\Gamma_j$) is relatively small, and thus, this jet may be
observable from a large solid angle.

\section{Discussion}

The centrifugal force due to rapid and differential rotation plays a
crucial role for supporting strong self-gravity of HMNS.  This
suggests that any HMNS will eventually collapse to a black hole after
a substantial loss of angular momentum by gravitational-wave
emission~\citep{STU2} and/or after a substantial angular momentum
transport inside it due to magnetic viscous effects~\citep{DLSSS} in a
realistic situation. The predicted lifetime is of order 10--100~ms.
After the collapse to a black hole, a system composed of a black hole
and compact accretion disk will be
formed~\citep{SDLSS,DLSSS,AEI,hotoke}. Then, the electromagnetic
radiation is likely to be emitted through the Blandford-Znajek
mechanism~\citep{BZ,MG-1,MG-2}. However, the lifetime of the accretion
disk will not be longer than $\sim 100$~ms because of the viscous
evolution.  Hence, the magnetic energy generation and mass ejection
are likely to continue only for a short time scale of order 100~ms
after the BNS merger.

However, the luminosity is quite high. If the kinetic energy of the
matter and/or electromagnetic energy are efficiently converted to an
electromagnetic signal, it will be observed even if the merger happens
at a distance of several hundred Mpc as discussed, e.g.,
in~\cite{NP11}.  Strong radio afterglow emission could be expected
when the jet propagates in the matter that may be ejected from neutron
stars and HMNS during the merger.  The ejected mass that can be
estimated as $\dot E_M\Delta T/c^2 \sim10^{-6}M_\odot B_{13}^2 R_6^3
\Omega_4 (\Delta T/0.1~{\rm s})$ could yield radioactive elements and
be observed like dim supernovae~\citep{macronova}.


In this article, we focus only on the HMNS with conservative magnetic
field strength $B \sim 10^{13}$~G (achieved by compression of ordinary
field strength $\sim 10^{12}$~G).  If the magnetic field strength were
as high as that of magnetar~\citep{magnetar}, i.e., $B \sim
10^{15}$~G, the electromagnetic luminosity would reach
$10^{51}$~ergs/s (see the bottom-left panel of Fig.~\ref{FIG2}).  This
value is as high as the luminosity of GRBs, and because the expected
time duration is less than 1~s, this is also a candidate model for
SGRB~\citep{SGRB}.  The canonical peak isotropic luminosity of SGRB is
$\sim 10^{51}$~ergs/s, which is consistent with this estimation,
assuming the jet solid angle as $\Omega_j\sim 0.1$ and the conversion
efficiency from the jet kinetic energy into the gamma rays as
$\eta\sim0.1$.  Hence, if one of neutron stars in BNS has a large
magnetic-field strength or the magnetic-field strength is
significantly amplified during the merger process or in the formed
HMNS, this may be observed as a SGRB.

\acknowledgments This work was supported by Grant-in-Aid for
Scientific Research (Nos. 19047004, 21340051, 21684014, 22244019,
22244030), by Grant-in-Aid for Young Scientists (B) (No. 22740178),
and by Grant-in-Aid for Scientific Research on Innovative Area
(No. 20105004), and HPCI Strategic Program of Japanese MEXT.

\end{document}